# Conductance modulation in Weyl semimetals with tilted energy dispersion without a band gap

Can Yesilyurt,[1,*] Seng Ghee Tan,[1,2] Gengchiau Liang,[1] and Mansoor B. A. Jalil[1,*]

[1]*Electrical and Computer Engineering, National University of Singapore, Singapore 117576, Republic of Singapore*

[2]*Data Storage Institute, Agency of Science, Technology and Research (A*Star), Singapore 138634, Republic of Singapore*

We investigate the tunneling conductance of Weyl semimetal with tilted energy dispersion by considering electron transmission through a p-n-p junction with one-dimensional electric and magnetic barrier. In the presence of both electric and magnetic barriers, we found that a large conductance gap can be produced by the aid of tilted energy dispersion without a band gap. The origin of this effect is the shift of the electron's wave-vector at barrier boundaries caused by i) the pseudo-magnetic field induced by electrical potential, i.e., a newly discovered feature that is only possible in the materials possessing tilted energy dispersion, ii) the real magnetic field induced by ferromagnetic layer deposited on the top of the system. We use realistic barrier structure applicable in current nanotechnology and analyze the temperature dependence of the tunneling conductance. The new approach presented here may resolve a major problem of possible transistor applications in topological semimetals, i.e., the absence of normal backscattering and gapless band structure.

## I. INTRODUCTION

In Dirac and Weyl semimetals, low-energy excitation of electrons exhibits linear dispersion denoting massless behavior of electrons with very high mobility [1]. The most well-known two-dimensional Dirac semimetal, graphene, was expected to be a promising candidate for transistor applications due to its high electron mobility. However, it has been revealed that the electrons are able to pass through high potential barriers without backscattering due to the chiral nature of the band structure [2]. While this feature opens new avenues for novel tunneling applications, it adversely affects the ability to modulate the electron transmission, since transmission is still possible regardless of barrier height, between the electron states in the conduction band and hole states in the valence band. Alternatively, the electron transmission may be controlled by changing the carrier concentration so as to tune the Fermi level into the band gap. However, the gapless band structure of monolayer graphene does not allow this feature. Besides, recent works have shows that these obstacles are not exclusive to graphene, but also applies to other Weyl semimetals with the conventional conical energy dispersion [3,4], as well as Weyl semimetals with tilted energy dispersion [5].

To overcome the conductance modulation issues addressed above, several solutions have been proposed based on band gap modulation. It is well-established that a graphene sheet can be structured into nanoribbons so as to induce an energy gap [6-9]. However, the resulting band gap is very sensitive to the fabrication process, and the carrier mobility is found to have decreased significantly. Other methods [10,11] based on the modulation of chemical structure of graphene also share the same drawbacks on the fabrication process and reduced mobility. Since the absence of backscattering is possible for only normally incident electrons in monolayer graphene, alternative geometry such as a saw-shape barrier geometry has been proposed to block the transmission of normally incident electrons in monolayer graphene [12], but this increases fabrication complexity. Here, we thus propose a distinct method to control electron transmission in Weyl semimetals, which requires a basic device configuration and without the need for an energy gap.

(*) email: can--yesilyurt@hotmail.com; elembaj@nus.edu.sg

The recently discovered Weyl semimetals can be considered as three-dimensional counterpart of graphene with higher carrier mobility as predicted and observed in various materials [13-16]. However, Weyl semimetals with conventional conical energy dispersion exhibit the same conductance modulation issues due to the chiral nature of Weyl fermions and the absence of a band gap [17]. A newly discovered type of Weyl semimetals [18,19] exhibits tilted energy dispersion of bulk states [20,21], where the Lorentz symmetry is broken. Recently, it has been shown that the application of electrical potential to a region of a Weyl semimetal with tilted energy dispersion causes a shift of electron's wave-vector (so called pseudo-magnetic field) as a consequence of broken Lorentz symmetry [5]. Here, we analyze the tunneling conductance of a Weyl system under the influence of both the pseudo-magnetic field induced by electrical potential, and the real magnetic field induced by ferromagnetic layer on the top of the system as illustrated in Fig. 1.

## II. METHOD

The electronic states in Weyl semimetals can be characterized as Weyl nodes [22] in bulk, and Fermi arc states [23] on the surface. A Weyl node cannot exist alone in the Brillouin zone, and must be paired with another Weyl node possessing the opposite chirality. Therefore, minimum two Weyl nodes must exist in the Brillouin zone, and they are connected to each other by the Fermi arc states. The robustness of a Weyl semimetal is crucial for possible tunneling applications and may be measured with the separation of Weyl nodes in $k$-space [24]. A general description of a Weyl fermion is given by the low energy Hamiltonian, i.e.,

$$H = V_0 + \sum_i \hbar \vec{k}_i \left( \sigma^i v_i + w_i \right), \tag{1}$$

where $\sigma^i$'s are Pauli matrices, $v_i$'s are velocities along the three directions, $w_i$'s are parameters denoting the "tilt" of the energy dispersion along the $i$th direction, and which are zero for conventional non-tilted Weyl fermion, while $V_0$ is the electrical potential applied to the system. Note that the sign of velocities $v_i$ determines the chirality of Weyl nodes. In this work, we neglect the contribution of Fermi arc states to the conduction, as this is shown to be negligible in a thin film system of the Dirac semimetal $Na_3Bi$ [25]. We also neglect inter-node scattering, as such scattering is suppressed in Weyl semimetals with well-separated Weyl nodes in their Brillouin zone, a required feature to ensure highly robust electronic states [24]. However, the transmission direction must be chosen carefully to avoid overlap between the Fermi surfaces corresponding to Weyl nodes of opposite chirality.

To demonstrate the conductance modulation of the p-n-p junction of a Weyl semimetal, we focus on electrons around a single node in an infinite Weyl semimetal system, as was the case in previous works [4,5,26]. Unlike previous works in the literature, which investigate conductance modulation by means of an energy gap opening, we aim to modulate the tunneling conductance by controlling the overlap between Fermi surfaces belong to different regions shown in Fig. 1. To shift the electrons momentum, we use two independent effects. The first one is pseudo-magnetic field [5] induced by an electrical potential in the central region of length $L$, which is described by $V_{(x)} = V_0[\Theta(x) - \Theta(x-L)]$. Assuming a Weyl system which is tilted only in the $y$-direction, the majority of the voltage dependent momentum shift must be along the $y$-direction. To generate such an electrostatic potential barrier, one needs to modulate the background carrier concentration of the whole system, as well as the carrier concentration within the central barrier region independently. This scheme may be achieved by using a back gate to tune the carrier concentration and hence, the Fermi level of the whole system, and simultaneously applying a top gate voltage to change the carrier concentration within only the central barrier region. This method has been commonly used in two-dimensional Dirac semimetals such as graphene, and has recently been demonstrated in the three-dimensional Dirac semimetal $Cd_3As_2$ [27]. The second effect is induced by a spatially localized real magnetic field at the boundaries of the central region, which induces a shift of electron's momentum in $k$-space, as will be explained below. Note that this shift differs crucially from the shift due to the pseudo-magnetic field. The latter shifts the Fermi surfaces but not the Weyl nodes, and hence the absence of backscattering for normally incident electrons remains. When the magnetization direction of the ferromagnetic layer is parallel to the transmission, the localized magnetic field can be approximately modeled as two asymmetric delta-function magnetic fields, i.e., $B_z = B_0 l_B [\delta(x) - \delta(x-L)]$, that corresponds to a piecewise constant magnetic gauge potential $\vec{A}_{B(z)} = B_0 l_B [\Theta(x) - \Theta(x-L)]\hat{y}$, where $l_B = \sqrt{\frac{\hbar}{|e|B_0}}$ [see Fig. 1 (d)]. This type of localized magnetic barrier can be generated by, either depositing ferromagnetic layer with in-plane magnetization on top of the barrier region [28-30] or by patterning two asymmetric ferromagnetic strips with out-of-plane magnetic anisotropy at the barrier boundaries [31-33]. The upper limit of the field strength $B_0$ depends on the saturation magnetization of the ferromagnetic material. For example, the saturation magnetization of magnetic barrier of up to 3.7 T has been achieved using dysprosium film [34].

The gauge potential $\vec{A}_{B(z)}$ modifies the Hamiltonian (Eq. 1) by shifting the wave-vector, i.e., $k_y = k_y + \frac{eA_{B(z)}}{\hbar}$. As can be seen from the Hamiltonian (Eq. 1), there is an additional shift due to the pseudo-magnetic field originating from the tilt strength $w_i$, as well as the applied electric potential $V_0$. Note that the application of the potential results in asymmetric shift of transverse wave-vector if the energy dispersion is tilted. The eigenenergies of the Hamiltonain (Eq. 1) are given by

$$\varepsilon_{\pm(\vec{k})} = \pm\hbar\left(\sqrt{k_x^2 v_x^2 + (k_y + \delta)^2 v_y^2 + k_z^2 v_z^2} + k_x w_x + (k_y + \delta)w_y + k_z w_z\right) + V_0 \qquad (2)$$

where $\delta = \frac{eA_{B(z)}}{\hbar}$, and the corresponding eigenstates are found by

$$\psi_{\pm} \equiv \frac{1}{\sqrt{2}} e^{i\vec{k}\vec{r}} \begin{pmatrix} 1 \\ e^{i\varphi} \sec\gamma(\pm 1 + \sin\gamma) \end{pmatrix} \equiv \begin{pmatrix} \psi_a \\ \psi_b^{\pm} \end{pmatrix}$$

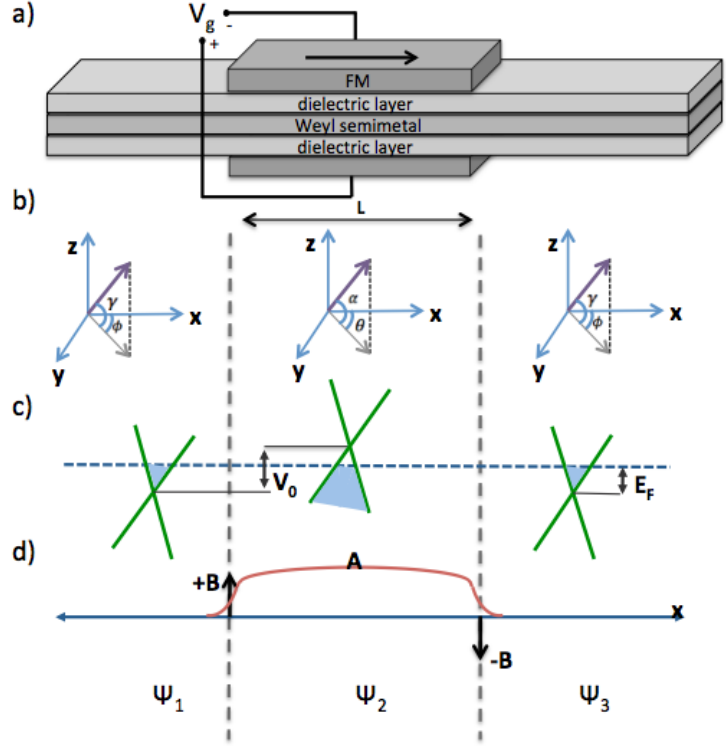

**Figure 1** The Weyl semimetal system with tilted dispersion on one direction. (a) shows the schematic illustration of proposed system and (b) shows the electron propagation angles outside and within the barrier region. (c) and (d) demonstrate the effect of applied electric potential and magnetic field induced by the ferromagnetic (FM) layer respectively.

The above wave functions describe the incident, propagated and transmitted electrons in the three regions shown in Fig. 1, denoted respectively with $\psi_1, \psi_2, \psi_3$. In the above, the angle $\gamma$ is between $\mathbf{k}$ and the $x$-$y$ plane, and $\varphi$ is the azimuthal angle with respect to the $x$-axis, as shown in Fig. 1(b), while transmission is along $x$-direction. Thus, outside the barrier region, the wavevectors can be written as $k_x = k_F \cos\gamma \cos\varphi$, $k_y = k_F \cos\gamma \sin\varphi$, $k_z = k_F \sin\gamma$. The transverse wave vectors $k_y$ and $k_z$ are conserved due to the translational invariance of the system along $y$- and $z$-directions. If the tilt is aligned along only the $y$-direction (and assuming $v_i = v_F$), the wave vector in the $x$-direction within the barrier is given by

$$q_{x(w_y,V_0)} = -\frac{1}{\hbar v_F}\sqrt{E_F^2 - 2E_F\left(V_0 + \hbar w_y(k_y+\delta)\right) + \left(V_0 + \hbar w_y(k_y+\delta)\right)^2 - \hbar^2 v_F^2\left(k_z^2 + (k_y+\delta)^2\right)} \quad (3)$$

In the barrier region, the angles of electron propagation $\theta = \tan^{-1}\left(\dfrac{k_y}{q_x}\right)$ and $\alpha = \tan^{-1}\left(\dfrac{k_z}{q_x}\cos\theta\right)$ can be found by considering the conservation of the transverse wave-vectors at the barrier interfaces.

The transmission probability of the system $T_{(\gamma,\varphi)}$ can then be calculated by matching the wave functions at the barrier interfaces. Finally, the total normalized conductance of the system is

$$G_{(E_F,V_0)} = G_0 \int dE \frac{-\partial f}{\partial E} \int_{-\pi/2}^{\pi/2} \int_{-\pi/2}^{\pi/2} d\gamma\, d\varphi\, \cos^2\gamma \cos\varphi\, T_{(\gamma,\varphi)} \qquad (4)$$

In the above equation, $f(E_F) = (1-\exp[(E-E_F)/(k_B T)])^{-1}$ is the Fermi distribution function, and $G_0 = \frac{e^2 S k_F^2 \upsilon}{4h\pi^2}$ is the quantum conductance in three-dimension derived by considering spin and valley degeneracy, where $\upsilon$ is the valley degeneracy, and $S$ the cross-sectional area of the system.

### III. RESULT AND DISCUSSION

We first analyze the conductance as a function of applied electrostatic potential $V_0$ at zero temperature for four combinations of Weyl systems concerning the existence of the tilt strength and magnetic field. We first consider a conventional (non-tilted) Weyl semimetal. As shown in Fig. 2 (a), the conductance as a function of applied potential exhibits a decreasing trend until $V_0 = E_F$, and then starts to increase with increasing $V_0$, which corresponds to the expected behavior in Dirac and Weyl semimetals owing to the conical energy dispersion. In the case of tilted Weyl system shown in Fig. 2 (b), the range of $V_0$ where the conductance is close to zero is even more restricted since the tilted Weyl fermion possesses larger density of states at low energies. Note that for the case where $w_i > v_i$ for any of the three directions, which is characteristic of type-II Weyl fermions, the density of states would be non-zero at intrinsic Fermi level. Thus having a tilted dispersion is disadvantageous from the viewpoint of conductance modulation (i.e. in suppressing the conductance) as far as Weyl semimetals are concerned. The situation is reversed upon application of the localized magnetic barrier. In the case of conventional (non-tilted) Weyl semimetal, the presence of the magnetic barrier increases the range of potential over which the conductance is close to zero, as shown in Fig. 2 (c). This is due to the larger mismatch of the Fermi surface of the central region with that of the source/drain as a result of aforementioned wavevector shift induced by $A_B$. However, the conductance is generally suppressed over the entire range of $V_0$ considered, and thus the system exhibits a smaller transconductance $\frac{\partial G}{\partial V_0}$, which is undesired in conductance modulation applications. Crucially, the application of magnetic barrier on the tilted Weyl system results not only in a large conductance gap, but also in a large $\frac{\partial G}{\partial V_0}$, as shown in Fig. 2 (d). This result can be understood as the consequence of a total mismatch of Fermi surfaces (with zero overlap), arising from the combination of two mechanisms that shift the Fermi surfaces, i.e., pseudo-magnetic field induced by the application of gate potential in a tilted Weyl semimetal, and transverse Lorentz displacement induced by the (real) magnetic barrier at the barrier boundaries.

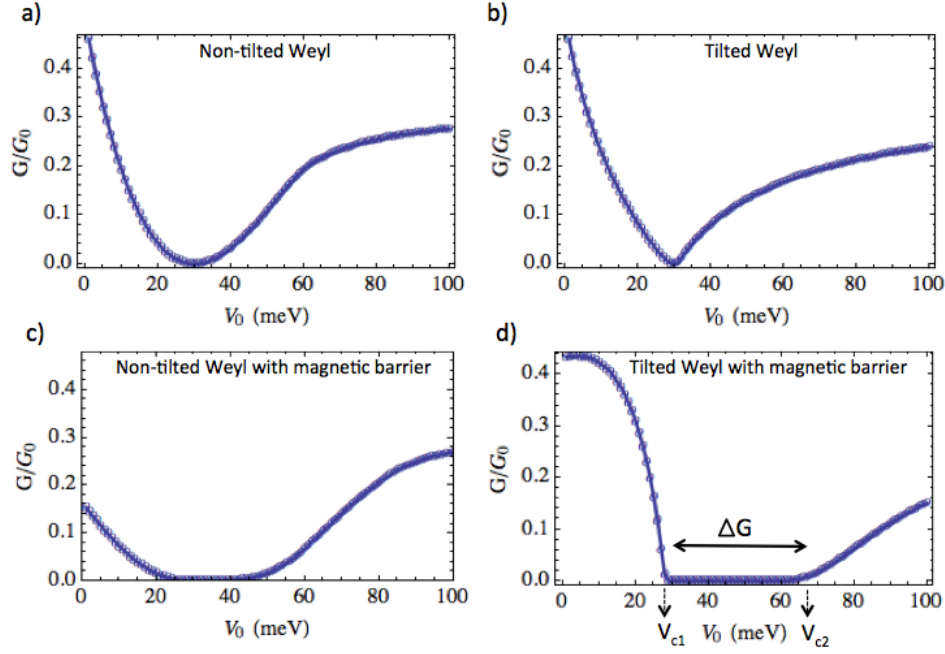

**Figure 2** Tunneling conductance as a function of gate voltage for different Weyl systems, (a) non-tilted Weyl system $(w_y/v_y = 0)$ in the absence of magnetic barrier $B_{0(z)} = 0$, (b) tilted Weyl system $(w_y/v_y = 0.9)$ in the absence of magnetic barrier $B_{0(z)} = 0$, (c) non-tilted Weyl system $(w_y/v_y = 0)$ in the presence of magnetic barrier $B_{0(z)} = -1.5\ T$, and (d) tilted Weyl system $(w_y/v_y = 0.9)$ in the presence of magnetic barrier $B_{0(z)} = -1.5\ T$. The Fermi energy $E_F = 30$ meV and barrier length $L = 900$ nm for all configurations.

Fig. 2 (d) numerically shows that the system possesses two critical electrical potential values $V_{c1}$ and $V_{c2}$, which define the conductance gap where the electrical conductance is almost zero. To find the analytical expression of these critical voltages, we consider the matching of the incident and propagated waves at the barrier interface $(x = 0)$. Intuitively, the conductance gap stems from the disjointedness of the range of occupied wave-vector $k_y$ values for the source and central regions. This arises due to the effect of both gauge potentials (from the pseudomagnetic and real magnetic fields) which cause a relative shift of the Fermi surfaces of these two regions in the $k_y$ direction. We focus on the extreme points of the occupied range of $k_y$ (for $k_x = 0$ and $k_z = 0$), inside $(k'_{y(\pm)})$ and outside $(k_{y(\pm)})$ the central barrier region. From the energy eigenvalue equation (2), we find that these are given by:

$$k_{y(\pm)} = \frac{\pm E_F}{\hbar v_F \pm \hbar w_y}, \tag{5}$$

$$k'_{y(\pm)} = \frac{\pm E_F \mp V_0 - \hbar v_F \delta \mp \hbar w_y \delta}{\hbar v_F \pm \hbar w_y}. \tag{6}$$

Each of the above equations represents two extreme $k_y$ points of the respective Fermi surfaces of the two regions, and so their equality $(k_{y(\pm)} = k'_{y(\mp)})$ represents the critical point at which the two surfaces become disjointed, leading to the aforementioned conductance gap. Thus, the value of $V_0$ that satisfies that equality actually corresponds to $V_{c1}$ which marks the start of the gap region. Explicitly, $V_{c1}$ is given by

$$V_{c1} = \frac{2E_F \hbar v_F \pm \hbar^2 v_F^2 \delta \mp \hbar^2 w_y^2 \delta}{\hbar v_F \pm \hbar w_y} \quad (7)$$

Note that there are two different $V_{c1}$ associated to the two disjointedness points of the Fermi surfaces, which depend on the direction of the shift due to the pseudo- and real magnetic fields. The value of the applied potential where the conductance again becomes non-zero, i.e., the critical value $V_{c2}$ is found by the direct equality of the extreme points $\left(k_{y(\pm)} = k'_{y(\pm)}\right)$, and given by $V_{c2} = \mp\left(\hbar v_F \pm \hbar w_y\right)\delta$. Substituting the expression for $\delta$ in the above equations, and assuming a negative shift in $k$-space, the conductance gap can be explicitly expressed in terms of the strength of the tilt and magnetic field barrier, as follows:

$$\Delta G = -\frac{2E_F \hbar v_F}{\hbar v_F + \hbar w_y} - 2v_F \eta \sqrt{\hbar q |B_{0(z)}|} \quad (8)$$

where $\eta$ is $(\pm 1)$ according to the direction of applied magnetic field. In Fig. 3, we analyze the conductance gap as a function of both the tilt strength $w_y$ and the magnetic field $B_{0(z)}$, which constitute the two main quantities that can be used to control the conductance gap. In general, the conductance gap can be widened by increasing the strength of both quantities.

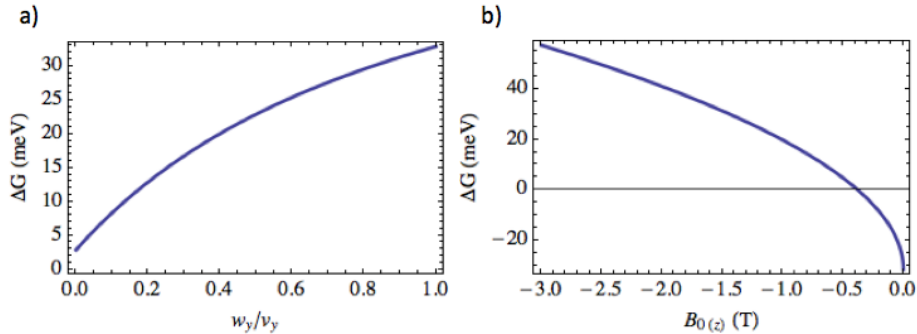

**Figure 3** Conductance gap $\Delta G$ shown in Fig. 2 (a) as a function of (a) the tilt strength of tilted Weyl fermions, where $B_{0(z)} = -1.5\ T$, (b) magnetic field strength induced by ferromagnetic layer, where $w_y = 0.9 v_y$. The Fermi energy $E_F = 30$ meV and barrier length $L = 900$ nm for both configurations.

As shown in Fig. 3, both the tilt strength and magnetic field can significantly modulate the conductance gap. Note that they must be combined with opposite signs in order to induce a conductance gap since the gauge potential induced by the electrical potential (pseudo-magnetic field) has opposite sign with reference to the tilt direction. Fig. 3 (b) reveals another interesting point that there is a critical magnetic field to obtain non-zero conductance gap, which directly depends on the extent of overlap of the Fermi surfaces in the $k_y$ direction, which is a function of Fermi energy. Therefore, one can further enhance the conductance gap by setting the Fermi level close to the Weyl node.

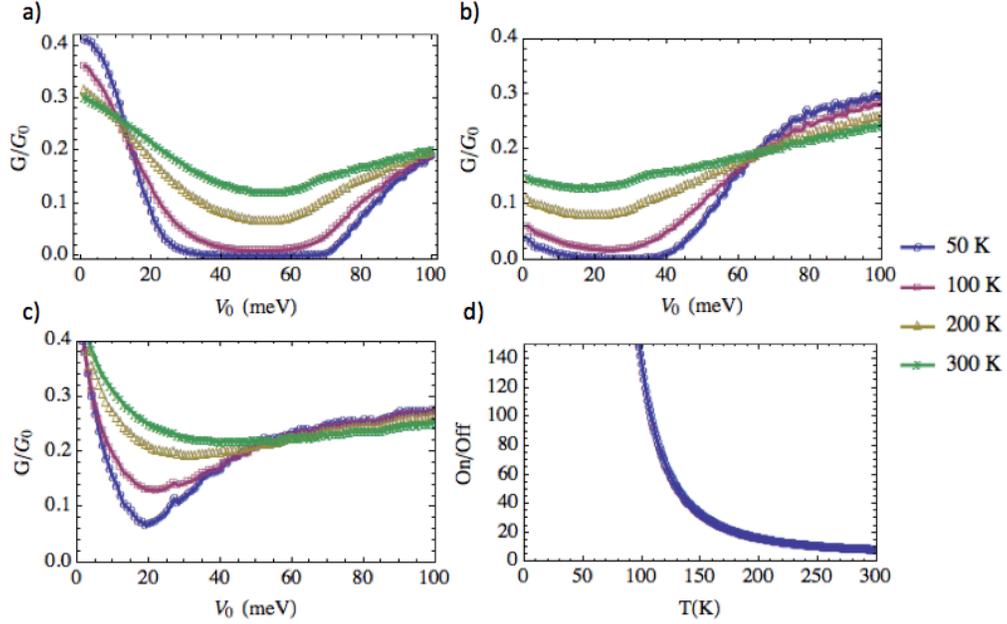

**Figure 4** Conductance as a function of gate voltage for different temperatures and configurations (a) tilted Weyl system $\left(w_y/v_y = 0.9\right)$ in the presence of magnetic barrier $B_{0(z)} = -2\ T$, (b) non-tilted Weyl system $\left(w_y/v_y = 0\right)$ in the presence of magnetic barrier $B_{0(z)} = -2\ T$, (c) tilted Weyl system $\left(w_y/v_y = 0.9\right)$ in the absence of magnetic barrier $B_{0(z)} = 0$, and (d) shows the temperature dependence of the on-off ratio for the configuration shown in (a), where on-voltage is 0 meV and off-voltage is 61 meV.

Finally, in Fig. 4, we analyze the effect of temperature on the conductance gap for the same set of configurations as that considered in Fig. 2. The conductance gap vanishes at room temperature for all configurations. The system is particular susceptible to temperature rise for the case where only the pseudomagnetic field is present [Fig. 4(c)], and the conductance gap disappears even at $T = 50$ K. However, the combination of both effects, i.e., pseudo- and real magnetic fields, yields a profile which is more robust to temperature rise, as shown in Fig. 4 (a). The on-off ratio of the system corresponding to Fig. 4 (a) shows extremely high values up to ~ 100 K, but degrades rapidly at higher temperatures approaching room temperature [see Fig. 4 (d)].

## IV. CONCLUSION

As a conclusion, large conductance gap can be induced by the application of electrical and magnetic potential barriers. We showed that the tilt strength of a Weyl semimetal can be utilized to enhance its conductance modulation. The effect of real magnetic and pseudo-magnetic fields in shifting the Fermi surface in *k*-space can be combined to yield a large conductance gap and a large transconductance gradient, two properties that can be very useful in electro-optic and conductance modulation applications. The present results may hold promise for resolving a major problem in Dirac and Weyl semimetals, i.e., their gapless band structure and suppression of normal backscattering, which preclude the formation of a conductance gap. The conductance gap remains distinct at $T = 50$ K, but does not survive up to room temperature. However, the room temperature on-off ratio may be improved by employing techniques such as a multiple barrier system, to enhance the $k$-space mismatch of the Fermi surfaces.

## ACKNOWLEDGEMENT

We acknowledge the financial support of MOE Tier II MOE2013-T2-2-125 (NUS Grant No. R-263-000-B98-112), and the National Research Foundation of Singapore under the CRP Programs "Next Generation Spin Torque Memories: From Fundamental Physics to Applications" NRF-CRP12-2013-01 (R-263-000-B30-281) and "Non-Volatile Magnetic Logic and Memory Integrated Circuit Devices" NRF-CRP9-2011-01 (R-263-000-A73-592).